\documentclass[
reprint,
superscriptaddress,
 amsmath,amssymb,
 aps,
pra,
]{revtex4-2}
\usepackage{makecell}
\usepackage{xcolor}
\usepackage{soul}
\usepackage{verbatim}
\usepackage{amsmath}
\usepackage{amssymb}
\usepackage{ulem}
\usepackage{xcolor}
\usepackage{braket}
\usepackage{caption}
\usepackage{subfigure}
\usepackage{graphicx}
\usepackage{bm}
\usepackage[misc,geometry]{ifsym}

\soulregister\cite7 
\soulregister\citep7 
\soulregister\citet7 
\soulregister\ref7 
\soulregister\pageref7 
\begin{document}


\title{Approaches of frequency-dependent squeezing for the low frequency detector of Einstein Telescope}
\author{Xingrui Peng}
\email{xxp381@student.bham.ac.uk}
\affiliation{Institute for Gravitational Wave Astronomy, School of Physics and Astronomy, University of Birmingham, Birmingham B15 2TT, United Kingdom}
\affiliation{School of Physical Science and Technology, Wuhan University, Wuhan, China 430072}
\author{Denis Martynov}
\affiliation{Institute for Gravitational Wave Astronomy, School of Physics and Astronomy, University of Birmingham, Birmingham B15 2TT, United Kingdom}
\author{Zonghong Zhu}
\affiliation{School of Physical Science and Technology, Wuhan University, Wuhan, China 430072}

\author{ Teng Zhang}
\email{tzhang@star.sr.bham.ac.uk}
\affiliation{Institute for Gravitational Wave Astronomy, School of Physics and Astronomy, University of Birmingham, Birmingham B15 2TT, United Kingdom}

\date{\today}

\begin{abstract}
The quantum noise in gravitational-wave detectors can be suppressed in a broadband by frequency-dependent squeezing. It usually requires one large scale filter cavity and even two, for example in the low frequency detector of Einstein Telescope, which is a detuned dual recycling Fabry-Perot Michelson interferometer. 
In this paper, we study the feasibility of replacing two filter cavities with a coupled-cavity, aiming to reduce the optical losses with less number of optics. It turns out this approach is only theoretically valid, however, the required parameters of the optics don't support practical implementation, which is consistent with the results in [Phys. Rev. D {\bf 101}, 082002
(2020)]. Furthermore, we investigate the viability of utilizing EPR squeezing to eliminate either one or two filter cavities in Einstein Telescope. It turns out EPR squeezing would allow to eliminate one filter cavity, and can potentially improve the detector sensitivity with the allowance of higher input squeezing level, benefiting from the longer length of the arm cavity which serves as one of the filter cavities for frequency-dependent squeezing. 
\end{abstract}

\maketitle

\section{Introduction\label{sec-introduction}}
Since the first gravitational-wave (GW) signal was detected in 2015, the network of Advanced LIGO\,\cite{LIGOScientific, PhysRevD102, PhysRevD93, PRL123, PhysRevLett.116.131103,Harry:2010zz}, Advanced Virgo\,\cite{VIRGO:2014yos,PhysRevLett.123.231108}, and KAGRA\,\cite{PhysRevD.88.043007,10.1093/ptep/ptaa125,Somiya_2012} have already detected nearly one hundred events\,\cite{GWTC3}, including binary black hole and neutron star coalescence. 
To explore further into the universe, a series of future detector plans aiming for a ten-fold overall sensitivity improvement were put into the discussion, such as Einstein Telescope (ET)\,\cite{Hild_2011} and Cosmic Explorer\,\cite{abbott2017exploring}. In these detectors, quantum noise limits their sensitivity in most of their frequency bands.
Quantum noise consists of radiation pressure noise and shot noise \cite{scalinglaw,quantummeasurement,rootyanbei}. Shot noise is due to the statistical fluctuation of the number of photons arriving at the photo detector in a unit of time, manifesting itself as fluctuations in the phase of the light and is important at high-frequency in the gravitational wave detectors. Radiation pressure noise comes from the test mass displacement raised by the fluctuating radiation pressure force from the light. It can be attributed to the amplitude fluctuation of the light. The radiation pressure noise is significant at low-frequency due to the mechanical dynamics of the test mass. The minimal sum of quantum shot noise and radiation pressure noise forms a lower bound given different power of the light, so called standard quantum limit (SQL)\,\cite{braginsky1995quantum}.

To overcome the standard quantum limit and improve the quantum noise limited sensitivity, various techniques were proposed. The frequency-dependent squeezing (FDS) has been applied and successfully suppressed the quantum noise in current detectors\,\cite{aasi2013enhanced, PhysRevLett.124.171102, PhysRevLett.131.041403}. In ET, there are two sets of detectors, which form a xylophone sensitivity. The high-power interferometer attains low shot noise and serves as the high-frequency detector; the low-power interferometer attains lower quantum radiation pressure noise and serves as the low-frequency detector. The low-frequency detector is a detuned interferometer in order to create optomechanical resonance at low frequency, which allows to overcome the SQL\,\cite{scalinglaw,Buonanno_2002,root}. In addition, FDS will be applied to further improve the overall quantum sensitivity. Therefore, it requires two filter cavities, which are responsible for compensating the phase rotation from the optical detuning and the optomechanical pondermotive effect, respectively. In practice, the two filter cavities in kilo-meter scale will entail substantial costs and bring more optical losses.
Alternative approaches of FDS are proposed in previous researches. One is to implement EPR squeezing, which was proposed the first time in\,\cite{ma2017proposal}, for an interferometer without detuning, and later on proposed for a detuned interferometer without signal recycling cavity (SRC)\,\cite{PhysRevD.96.062003}. The EPR squeezing scheme was experimentally demonstrated in \,\cite{sudbeck2020demonstration,yap2020generation,PhysRevResearch.3.04309}. Recently, a proposed FDS scheme using quantum teleportation (QT) technology can eliminate both filter cavities in a detuned dual-recycling Fabry-Perot Michelson interferometer (DRFPMI)\,\cite{nishino2024frequency}. Another idea is to use a coupled filter cavity to replace two individual filter cavities. It was proposed first in\,\cite{ET2filter}, while only numerical result was given.

In this paper, following the previous study, we systematically study the feasibility of alternative approaches to conduct FDS for a detuned DRFPMI and compare them with the conventional approach and the new QT approach.
In section\,\ref{sec-roots}, we review the input-output relation of the interferometer, based on which the required squeezing rotation angle and parameters of filter cavities can be derived. In section\,\ref{sec-2filtercoupled}, we analytically study the equivalence of two filter cavities and a coupled-cavity, which inspires the motivation of taking the coupled SRC-arm cavity as the filter cavity for EPR squeezing.
In section\,\ref{sec-EPR}, 
we conduct the research of EPR squeezing to eliminate both required filter cavities, and show that this scheme is infeasible given the constraints of the interferometer parameters.
In section\,\ref{sec-EPRparameters}, we propose to use the EPR scheme to replace one of the two filter cavities, the filter cavity with narrower bandwidth, more specifically, and compare the resulting sensitivities with those of the conventional approach and QT approach after including the optical losses.

\section{Frequency-dependent squeezing via filter cavities\label{sec-roots}}
In the two-photon formalism\,\cite{PhysRevA.31.3068,PhysRevA.31.3093,PhysRevA.72.013818}, the input-output relation of the interferometer can be generally written as:
\begin{equation}\label{eq-generalinputoutput}
\hat{o}=\mathbb{T} \hat{i}+\mathbf{R}\hat{x}\,,
\end{equation}
where $\hat{o}$ and $\hat{i}$ are the output and input fields of the interferometer at the dark port. The expression of transfer matrix $\mathbb{T}=\begin{bmatrix}T_{11}(\Omega) & T_{12}(\Omega)\\T_{21}(\Omega) & T_{22}(\Omega) 
\end{bmatrix}$ is given as Eq.(2.20)  in \cite{two-photon1}. $\mathbf{R}$ is the response vector of the interferometer to the differential displacement of the test mass, $\hat{x}$. The quantum noise spectral density of the interferometer can be calculated in the general form as\,\cite{quantummeasurement}:
\begin{equation}\label{homodetection}
S=\frac{\mathbf{H}\mathbb{T}\mathbb{P}
\begin{bmatrix}e^{2r}& 0\\ 0 &e^{-2r}
\end{bmatrix}\mathbb{P}^{\dagger}\mathbb{T}^{\dagger}\mathbf{H}^{\rm T}}{|\mathbf{H}\mathbf{R}|^2}\,,
\end{equation}
where $\mathbf{H}=\begin{bmatrix}\cos\zeta &\sin \zeta
\end{bmatrix}$ is the homodyne detection vector with homodyne angle $\zeta$; $\mathbb{P}=\begin{bmatrix}\cos \theta  & \sin \theta \\ -\sin \theta & \cos  \theta \end{bmatrix}$ is the rotation matrix with rotation angle $\theta$, usually provided by the filter cavities; $r$ is the squeezing factor with positive sign meaning phase quadrature squeezed.
In order to make Eq.\,\ref{homodetection} minimum, the rotation angle $\theta$ should satisfy:
\begin{equation}\label{rotationangle}
    \tan\theta=-\frac{T_{11}(\Omega)\text{cos}\zeta+T_{21}(\Omega)\text{sin}\zeta}{T_{12}(\Omega)\text{cos}\zeta+T_{22}(\Omega)\text{sin}\zeta}\, .
\end{equation}
The Eq.\,\ref{rotationangle} can be expressed in a general format as follows:
\begin{equation}
    \text{tan}\theta(\Omega)=\frac{\sum_{k=0}^{n}B_k\Omega^{2k}}{\sum_{k=0}^{n}A_k\Omega^{2k}}\, ,
\end{equation}
where $A_k$ and $B_k$ are real constants with $A_k^2+B_k^2>0$. As detailed in the Appendix.A of \cite{rootyanbei}, 
the bandwidth and detuning of filter cavities for FDS can be derived by calculating the complex roots of the equation:
\begin{equation}\label{eq-rootequation}
    \sum_{k=0}^{n}(A_{k}+iB_{k})\Omega^{2k}=0\, ,
\end{equation}
with the roots, $\Omega_j=\gamma_j+i\delta\omega_j,\, j=1,2,...,2k$, where $\gamma_j$ and $\delta\omega_j$ stand for the bandwidth and detuning of the filter cavities. Taking the parameters of ET as the example, which are shown in Table.\,\ref{tab4}, the parameters of the two required filter cavities can be calculated as:
\begin{subequations}\label{eq-2filterparameters}
    \begin{equation}
    \frac{\gamma_1}{2\pi}=4.26 \,\text{Hz},\quad \frac{\delta\omega_1}{2\pi}=19.51 \,\text{Hz}\, ,
    \end{equation}
    \begin{equation}
    \frac{\gamma_2}{2\pi}=1.65 \,\text{Hz}, \quad\frac{\delta\omega_2}{2\pi}=-7.65 \,\text{Hz}\, .
    \end{equation}
\end{subequations}

\begin{table}[!t]
\centering
\begin{tabular}{c|c|c}
\hline
\hline
Parameter & Physical meaning & Value  \\ \hline
$M$  & mass of the test mass &  211 kg  \\ 
$I$  & power of the laser &  63 W  \\ 
$L_{\text{SRC}}$  & SRC length &  152 m/86 m  \\  
$L_{\text{1,2}}$  & filter cavity length &  $10^{3}$ m  \\  
$L_{\text{arm}}$  & arm length & 10000.3 m/10000.2 m  \\ 
$\Delta$  &  frequency detuning (idler) & 1.27 MHz/2.25 MHz  \\
$T_{\text{SRM}}$  & power transmissivity of SRM & 20\%  \\
$T_{\text{ITM}}$  & power transmissivity of ITM & 0.7\%  \\
$\varphi_{\rm SRC}$  & round-trip detuning of SRC & 0.75 rad  \\
$\zeta_{s}$    & homodyne angle (signal) &$\frac{\pi}{2}$  \\ 
$\zeta_{i}$ &homodyne angle (idler) & $\frac{\pi}{2}$+0.1  \\
$r$ &squeezing factor & 1.15/1.73  \\
$\epsilon_\text{i}$  & input loss &  4\%  \\  
$\epsilon_\text{r}$  & readout loss & 3\%   \\ 
$\epsilon_{\text{SRC}}$  &  SRC loss & 1000 ppm \\ 
$\epsilon_{\text{arm}}$    & arm cavity loss &45 ppm  \\ 
$\epsilon_{\text{f}}$ &filter cavity loss & 20 ppm \\
$\Delta L_{\rm f}$ & filter cavity length deviation & $10^{-12}$ m\\
\hline
\hline
\end{tabular}
\caption{The table shows the parameters of the interferometer}
\label{tab4}
\end{table}

\section{Equivalence between two filter cavities and a coupled-cavity\label{sec-2filtercoupled}}
\begin{figure}[!t]
\includegraphics[width=0.95\columnwidth]{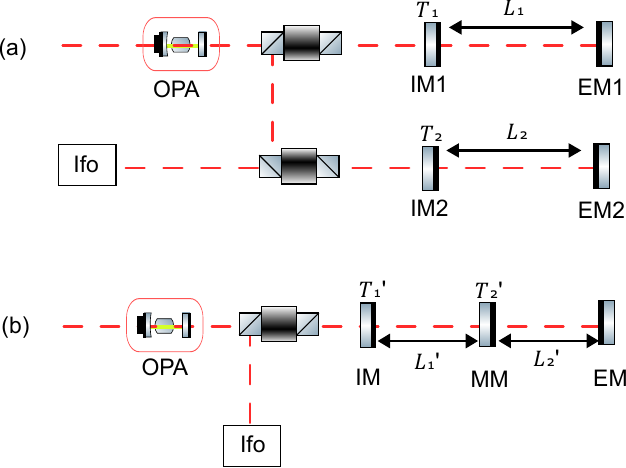}
\caption{(a) The upper figure shows the set-up of the two filter cavities scheme; (b) The lower figure shows the set-up of the coupled-cavity scheme. We use the optical parametric amplifier (OPA) to produce the squeezed light, then inject the squeezed light into either two filter cavities or the coupled-cavity. After the beam is reflected from the filter cavity, we feed the beam into the interferometer (Ifo) from the dark port}
\label{fig-2filter}
\end{figure}
In this section, we investigate the equivalence between two sequential filter cavities and a three-mirror coupled-cavity to achieve the FDS.
\cite{ET2filter} studies the two main advantages of the coupled-cavity: (1) compared to the two filter cavities, the coupled-cavity doesn't need some auxiliary optics such as the Faraday isolator, without which the overall optical loss in the input squeezing path can be reduced; (2) in the case of using the coupled-cavity, the two cavities can be arranged sequentially in one long vacuum system, whereas the two separated filter cavities require a shorter but wider vacuum space.

As detailed in Appendix.\,\ref{app-2filter}, concerning only the upper sideband of light, the input-output transfer functions are:
\begin{subequations}\label{eq-inputoutput}
\begin{equation}\label{2flterapp1}
f(\Omega)=\frac{\frac{1}{4}T_1T_2-\phi_{1}(\Omega)\phi_{2}(\Omega)+\frac{i}{2}(T_2\phi_{1}(\Omega)+T_1\phi_{2}(\Omega))}{\frac{1}{4}T_1T_2-\phi_{1}(\Omega)\phi_{2}(\Omega)-\frac{i}{2}(T_2\phi_{1}(\Omega)+T_1\phi_{2}(\Omega))}\, ,
\end{equation}
\begin{equation}\label{couplapp1}f'(\Omega)=\frac{T_2'-\phi_{1}'(\Omega)\phi_{2}'(\Omega)+\frac{i}{2}T_1'\phi_{2}'(\Omega)}{T_2'-\phi_{1}'(\Omega)\phi_{2}'(\Omega)-\frac{i}{2}T_1'\phi_{2}'(\Omega)}\, ,
\end{equation}
\end{subequations}
where $f(\Omega)$ is the input-output relation of the two filter cavities, $\phi_{1,2}(\Omega)=2(\delta\omega_{1,2} +\Omega)\frac{L_{1,2}}{c}$ are the round trip phases the light acquires in corresponding cavities, $\Omega$ is the sideband frequency, $c$ is the speed of light, $L_{1,2}$ are the cavity lengths of corresponding cavities, and $T_{1,2}$ are the power transmissivity of the two input mirrors (IM1,IM2) as shown in (a) of Fig\,\ref{fig-2filter}, and $f'(\Omega)$ is the input-output relation of the coupled-cavity, $\phi_{1,2}'(\Omega)=2(\delta\omega_{1,2}' +\Omega)\frac{L_{1,2}'}{c}$ are the round trip phases the light acquires, $\delta\omega_{1,2}'$ are the detunings of the corresponding cavities, $T_1'$ is the power transmissivity of the input mirror (IM), $T_2'$ is the power transmissivity of the middle mirror (MM), $L_{1,2}'$ are the cavity lengths as shown in (b) of Fig\,\ref{fig-2filter}. 

By comparing the real and imaginary parts of the numerator or denominator of Eq.\,\ref{2flterapp1} and Eq.\,\ref{couplapp1} and setting them identical, we find below relations of the parameters:
\begin{subequations}\label{2fltertocoupling}
    \begin{equation}
    L_{1}'L_{2}'=L_{1}L_{2}\,,
    \end{equation}
    \begin{equation}
    \gamma_{1}'=\gamma_{1}+\gamma_{2}\,,
    \end{equation}
    \begin{equation}
    \delta\omega_{1}'=\frac{\gamma_{1}}{\gamma_{1}+\gamma_{2}}\delta\omega_{1}+\frac{\gamma_{2}}{\gamma_{1}+\gamma_{2}}\delta\omega_{2}\,,
    \end{equation}
    \begin{equation}
    \delta\omega_{2}'=\frac{\gamma_{2}}{\gamma_{1}+\gamma_{2}}\delta\omega_{1}+\frac{\gamma_{1}}{\gamma_{1}+\gamma_{2}}\delta\omega_{2}\,,
    \end{equation}
    \begin{equation}
    \omega_{s}=\sqrt{[1+(\frac{\delta\omega_{1}-\delta\omega_{2}}{\gamma_{1}+\gamma_{2}})^2]\gamma_{1}\gamma_{2}}\,,
    \end{equation}
\end{subequations}
where $\gamma_{1}'=\frac{cT_{1}'}{4L_{1}'}$ is the half-bandwidth of the first cavity in the coupled-cavity, $\omega_{s}$ is the splitting frequency of the two resonances of the coupled-cavity: 
\begin{equation}\omega_s=\frac{c\,\sqrt{T_{\text{2}}'}}{2\sqrt{L_1'L_2'}}\, .
\end{equation}

\begin{figure}[!t]
\includegraphics[width=\columnwidth]{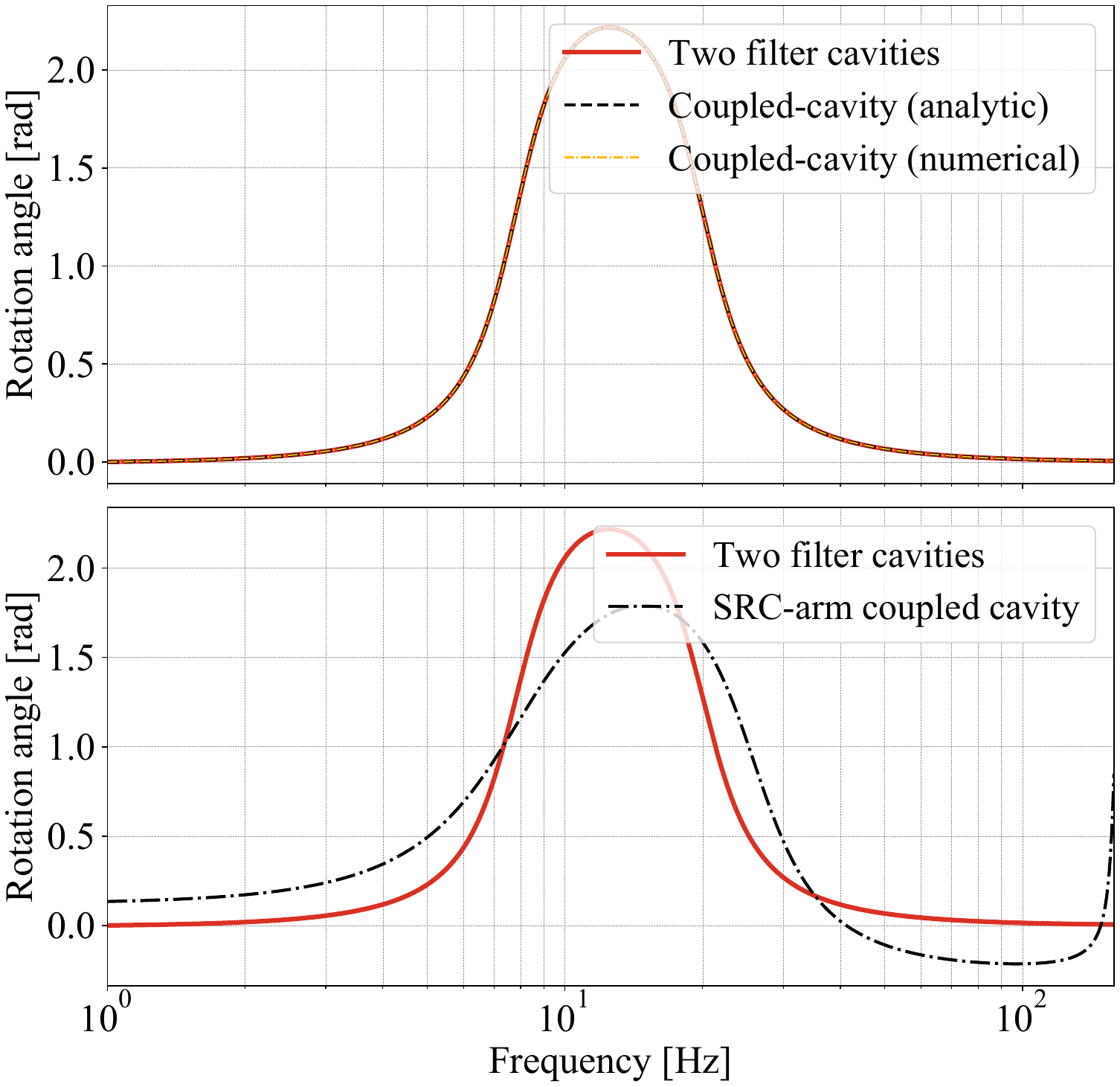}
\caption{(a) The upper plot shows the calculated rotation angles using the parameters of the coupled-cavity derived from both the analytical method and numerical nonlinear least square method approaches. They agree with the rotation angle provided by the two-cavity set-up.
(b) The lower plot shows the rotation angles of the SRC-arm coupled-cavity for the idler beam. The rotation angles are calculated with the parameters of the SRC-arm coupled-cavity fitted by Python and don't agree with the rotation angle provided by the two-cavity set-up}
\label{fig-rotation}
\end{figure}

By bringing the parameters in Eq.\,\ref{eq-2filterparameters} into Eq.\,\ref{2fltertocoupling}, we can calculate the parameters of the coupled-cavity. We also search the coupled-cavity parameters numerically by nonlinear least square method to match Eq.\,\ref{2flterapp1} and Eq.\,\ref{couplapp1}, and encouragingly, the calculated rotation angles using both methods are consistent, as digitally shown in Table.\,\ref{tab1} and visually shown in (a) of Fig.\,\ref{fig-rotation}. It proves the equivalence of the two schemes. However, the calculated transmissivity of the MM is:
\begin{equation}
T_2'\simeq 2.7\times10^{-7}\,,
\end{equation}
which is unrealistically small considering the losses of the optics and the manufacturing of the optical coatings.

\begin{table}[!t]
\centering
\begin{tabular}{c|c|c|cc}
\hline
\hline
Parameter & Value & Parameter & \multicolumn{2}{c}{\makecell[c]{Value\\ Analytic $~$ Numerical}} \\ 
\hline
$\delta\omega_1$  & 19.51 $\rm Hz$ &$\delta\omega'_1$  & 11.94 $\rm Hz$ &11.93 $\rm Hz$ \\  
$\delta\omega_2$  & -7.65 $\rm Hz$&$\delta\omega'_2$  & -0.08 $\rm Hz$&-0.07 $\rm Hz$ \\ 
$\gamma_1$  & 4.26 $\rm Hz$ &$\gamma'_1$  & 5.90 $\rm Hz$& 5.91 $\rm Hz$\\ 
$\gamma_2$  & 1.65 $\rm Hz$ &$\delta\omega_{s}'$  & 12.46 $\rm Hz$ &12.47 $\rm Hz$ \\ 
$L_1$  & 1000 $\rm m$&$L_1'$  & 1000 $\rm m$& 1000 $\rm m$  \\ 
$L_2$  & 1000 $\rm m$&$L_2'$  & 1000 $\rm m$& 1000 $\rm m$ \\ 
\hline
\hline
\end{tabular}
\caption{The table shows the values of the parameters of the two filter cavities and the coupled-cavity. The parameters on the left are the calculated parameters of two filter cavities. The parameters on the right are the calculated and the fitted parameters of the coupled-cavity}
\label{tab1}
\end{table}

\section{Replacing two filter cavities using EPR-entanglement scheme\label{sec-EPR}}
Enlightened by the study in the last section, in this section, we discuss whether the coupled SRC-arm cavity can perform as the filter cavity in the EPR squeezing scheme. The principle of the EPR scheme is studied adequately in \cite{EPR,NaturePhotonics}. Here, we review the core steps. Firstly, the squeezer is pumped at frequency $\omega_p=2\omega_0+\Delta$, where $\omega_0$ is the carrier frequency of the interferometer, such that the signal field $\hat{a}$ with frequency $\omega_0$, and idler field $\hat{b}$ with frequency $\omega_0+\Delta$ are entangled with each other in the EPR way. Secondly, the two fields are injected into the interferometer from the dark port, thereby the quadratures of the idler field would rotate by seeing the interferometer as the filter cavity. The quadrature rotation of the idler field is required to be inverse to the rotation of the signal field due to optical detuning and ponderomotive effect. Finally, the signal and idler beams are separately detected with homodyne detection, and then their photocurrents are combined with an optimal filter.

Note that in \cite{EPR}, the phase gained by sideband at frequency $\Omega$ in the SRC is ignored, since the length of SRC is much smaller than that of the arm cavity and $\Omega \ll \Delta$. Therefore, the SRC can be treated as a compound mirror that forms a single filter cavity together with the end test mass (ETM) of the arm cavity. Here we precisely analyze the system and treat the coupled SRC-arm cavity as a three-mirror cavity.
As detailed in the Appendix.\,\ref{app-2filter}, we derive the parameter relations between the coupled SRC-arm cavity and the coupled-cavity shown in (b) of Fig.\,\ref{fig-2filter} as: 
\begin{subequations}\label{eq-constraints}
    \begin{equation}
L_{\rm SRC}L_{\rm arm}=L_{1}'L_{2}'\,,
    \end{equation}
    \begin{equation}
    \gamma_{\rm SRC}= \gamma_{1}'\,,
    \end{equation}
    \begin{equation}
\delta\omega_{\rm SRC}=\delta\omega_{1}'-\delta\omega_{2}'\,,
    \end{equation}
    \begin{equation}
    \delta\omega_{\rm arm}=0,
    \end{equation}
    \begin{equation}
    \Delta=\delta\omega_{2}'\,,
    \end{equation}
    \begin{equation}\label{eq-Tconstraints}
    \omega_{s}^{\text{SRC-arm}}=\omega_{s}\,,
    \end{equation}
\end{subequations}
where $L_{\rm SRC}$ is the length of the SRC, $L_{\rm arm}$ is the length of the arm cavity, $\gamma_{\rm SRC}=\frac{cT_{\rm SRM}}{4L_{\rm SRC}}$ is the half-bandwidth of the SRC, $T_{\rm SRM}$ is the power transmissivity of the signal recycling mirror (SRM), $\delta\omega_{\rm SRC}$ is the detuning of the SRC to the signal beam, $\delta\omega_{\rm arm}$ is the detuning of the arm cavity to the signal beam, $\Delta$ is the frequency detuning of the idler beam, $\omega_{s}^{\text{SRC-arm}}$ is the splitting frequency:
\begin{equation}
    \omega_{s}^{\text{SRC-arm}}=\frac{c \sqrt{T_{\rm ITM}}}{2\sqrt{L_{\rm SRC}L_{\rm arm}}}\, .
\end{equation}
Here $T_{\rm ITM}$ is the power transmissivity of the input test mirror (ITM).

Eq.\,\ref{eq-constraints} shows there are some degrees of freedom we can adjust, \textit{i.e.} the SRC and arm cavity length. However, we find that Eq.\,\ref{eq-Tconstraints} strictly requires the transmissivity of the ITM should be $2.7\times10^{-7}$ but the transmissivity of ITM of ET is $7\times10^{-3}$.  The numeric result also suggests that the rotation angle provided by the interferometer isn't consistent with the rotation angle provided by two filter cavities, as shown in (b) of Fig.\,\ref{fig-rotation}. As a conclusion, the EPR scheme of employing the SRC-arm cavity to eliminate both filter cavities of ET is infeasible.

\section{Replacing one filter cavity using EPR-entanglement scheme\label{sec-EPRparameters}}
\begin{figure}
\includegraphics[width=1\columnwidth]{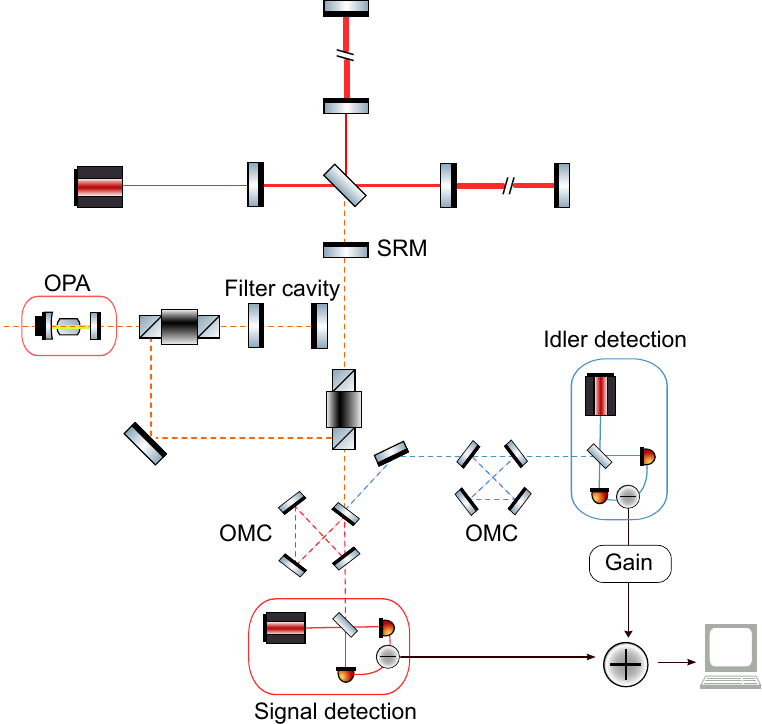}
\caption{The figure shows the setup for eliminating one required filter cavity by using the EPR squeezing scheme. At first, We pump the OPA to produce a signal beam with frequency $\omega_0$ and an idler beam with frequency $\omega_0+\Delta$, which are entangled with each other in an EPR way. Then we inject them into a filter cavity, which is anti-resonant to the signal beam. After being reflected from the filter cavity, we feed them into the interferometer from the SRM. Finally, after two beams interact with the interferometer and are reflected from the SRM, we use OMC to separate these two beams from each other to measure them separately and then combine the measurement with an optimal filter}
\label{fig-ETEPR}
\end{figure}

Using the EPR scheme to eliminate both required filter cavities has been proven to be infeasible in section.\,\ref{sec-EPR}. In this section, we step back and utilize the EPR squeezing to eliminate one of the required filter cavities, as shown in Fig.~\ref{fig-ETEPR}.
\subsection{Parameter choice}
Here the remaining filter cavity is only expected to provide quadrature rotation to the idler beam. In order to avoid undesired rotation to the signal beam, the signal beam needs to be kept either on resonance or anti-resonance in the filter cavity. The latter option which can give less optical losses is chosen as the setup in the paper. The idler beam is detuned from the filter cavity resonance by $\delta \omega_1$, therefore the frequency difference between idler beam and signal beam, $\Delta$ needs to satisfy:
\begin{equation}\label{eq-EPR-filter}
\Delta=\delta\omega_1+(2 n_1+1)\pi\times \frac{c}{2L_{1}},n_1\in\mathbf{Z}\, ,
 \end{equation}
The interferometer will serve as the second filter cavity equivalently for the idler beam owning the bandwidth $\gamma_2$ and detuning $\delta \omega_2$. The interferometer bandwidth reads:
  \begin{equation}\label{eq:gamma2}
    \gamma_2=\frac{c |\mathcal{T}_{\rm SRC}|^2}{4L_\text{arm}}\,,
\end{equation}
where $\mathcal{T}_{\rm SRC}$ is the effective amplitude transmissivity of SRC and can be calculated as:
\begin{equation}
\mathcal{T}_{\rm SRC}=\frac{\sqrt{T_{\text{ITM}}T_{\text{SRM}}}e^{i\frac{\phi_{\rm SRC}}{2}}}{1-\sqrt{R_{\text{ITM}}R_{\text{SRM}}}e^{2i\phi_{\text{SRC}}}}\,.
\end{equation}
Here $\phi_{\text{SRC}}=2(\Delta \frac{L_{\rm SRC}}{c}+\varphi_{\text{SRC}})$ is the round trip phase the idler beam acquires in the SRC, $\varphi_{\text{SRC}}$  is the constant detuning of the SRC, $R_{\rm ITM, SRM}$ is the power reflectivity of the ITM and SRM.
The effective detuning, $\delta \omega_2$, needs to satisfy:
\begin{equation}\label{eq-resonantconstraint}
2(\delta\omega_2+\Delta)\frac{L_{\text{arm}}}{c}+\text{Arg}[\mathcal{R}_{\rm SRC}]=2n_2\pi,n_2\in\mathbf{Z}\,,
\end{equation}
where $\mathcal{R}_{\rm SRC}$ is the effective amplitude reflectivity of SRC when the beam is injected from the ITM: 
\begin{equation}
\mathcal{R}_{\rm SRC}=\frac{\sqrt{R_{\text{ITM}}}-\sqrt{R_{\text{SRM}}}e^{i\phi_{\text{SRC}}}}{1-\sqrt{R_{\text{ITM}}R_{\text{SRM}}}e^{i\phi_{\text{SRC}}}}\, .
\end{equation}

According to Eqs.\,\ref{eq-EPR-filter},\ref{eq:gamma2},\ref{eq-resonantconstraint} and the parameters of ET, the detuning $\Delta$, SRC length and fine-tuning of the arm cavity length can be derived as shown in Table.~\ref{tab4}. Here we constrain the SRC length below 200\,m and chose two values around 80\,m and 150\,m as examples.

\subsection{Detector sensitivity}
We denote the output of the signal beam and the idler beams as  $\hat{A}$ and $\hat{B}$, and the transfer functions from input fields ($\hat{a}, \hat{b}$) to the output fields ($\hat{A}, \hat{B}$) are derived in Appendix.~\ref{app-loss}. At the homodyne detection, their phase quadratures are measured (denoted by subscript 2), and the combined measurement after filtering $\hat{B}_2$ by an optimal filter '$g$' can be written as:
\begin{subequations}
    \begin{equation}
\hat{A}_g=\hat{A}_2+g\hat{B}_2, \quad  g= -\frac{S_{\hat{A}_2\hat{B}_2}}{S_{\hat{B}_2\hat{B}_2}}\, .
\end{equation}
The power spectrum density (PSD) of $\hat{A}_g$ can be calculated as:
\begin{equation}
\begin{aligned}
    S_{\hat{A}_g\hat{A}_g}=&S_{\hat{A}_2\hat{A}_2}+|g|^2S_{\hat{B}_2\hat{B}_2}+g^*S_{\hat{A}_2\hat{B}_2}+gS_{\hat{B}_2\hat{A}_2}\, ,
\end{aligned}
\end{equation}
\end{subequations}
where $S_{{\hat{A}_2}{\hat{A}_2}}, \, S_{\hat{B}_2\hat{B}_2}$ are the PSD of $\hat{A}_2, \hat{B}_2$, and $S_{\hat{A}_2\hat{B}_2},\, S_{\hat{B}_2\hat{A}_2}$ are their cross PSD. 

\begin{figure}[!t]
\includegraphics[width=1\columnwidth]{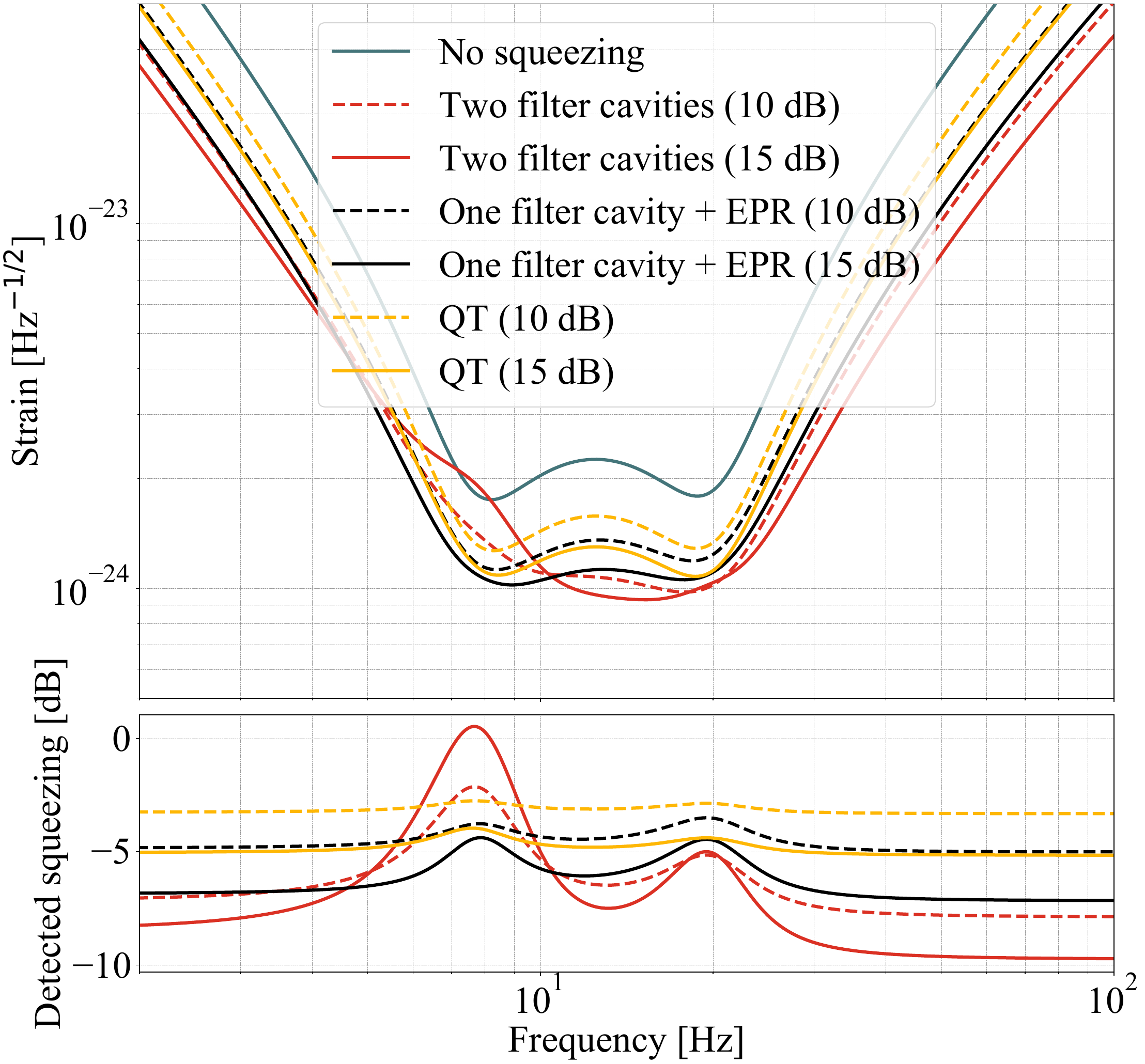}
\caption{(a) The upper figure shows the comparison of the quantum sensitivity of ET low frequency detector using different FDS schemes; (b) The lower figure shows the detected squeezing level from different FDS schemes} 
\label{fig-EPRsensitivity}
\end{figure}

The sensitivities are calculated with 10\,dB and 15\,dB input squeezing, as shown in Fig.~\ref{fig-EPRsensitivity}. As a comparison, we also show the result of the conventional FDS scheme with two filter cavities and the recently studied QT squeezing technology\,\cite{nishino2024frequency}. 

In general, the EPR scheme incurs an intrinsic $\sim$ 3 dB penalty due to the lack of response of the idler beam to the GW signal. Besides, it is susceptible to double optical losses, which cumulatively arise from two optical paths. However, leveraging the arm cavity to replace a filter cavity enables us to lengthen the effective filter cavity, hence reduces the effective filter cavity loss and phase noise, assessed through the term per unit length. 

In the FDS scheme using detuned filter cavities, the filter cavity loss will introduce a magnitude imbalance between the upper and lower sideband, which causes the so called dephasing effect, in particular around the detuning frequency. Such an effect couples the undesired anti-squeezing to the phase quadrature\,\cite{PhysRevD.104.062006, PhysRevD.90.062006}. Therefore, around 8\,Hz, where the detuning frequency of the filter cavity with narrower bandwidth locates and the optical optomechanical resonance strengthens the ponder-motive squeezing, the sensitivity of the two filter cavities scheme becomes worse after increasing the input squeezing from 10\,dB to 15\,dB.  In contrast, in the EPR squeezing setup, we replace the narrower bandwidth filter cavity with a longer arm cavity, such a dephasing contamination gets moderated, which allows the 15\,dB scheme to show superiority. Given the constraint on physical lengths of the external filter cavities to 1\,km, the conventional FDS prefers lower input squeezing as\,10\,dB, while the EPR squeezing could in principle allow higher input squeezing for a better performance.


The QT scheme replaces both the filter cavities with the arm cavity, effectively enhances the lengths of both filter cavities. Therefore, it shows almost constant sensitivity enhancement by increasing the input squeezing. However, it involves three beams at the detection with only one beam sensing the gravitational wave signal, therefore, the inherent squeezing penalty is $\sim$ 4.8 dB. In the meanwhile, there exist threefold optical losses\,\cite{nishino2024frequency}. It gives overall worse sensitivity compared with the EPR scheme, while also shows superiority around 8\,Hz compared with the conventional FDS with 15\,dB input squeezing.  

\begin{figure}[!t]
\includegraphics[width=1\columnwidth]{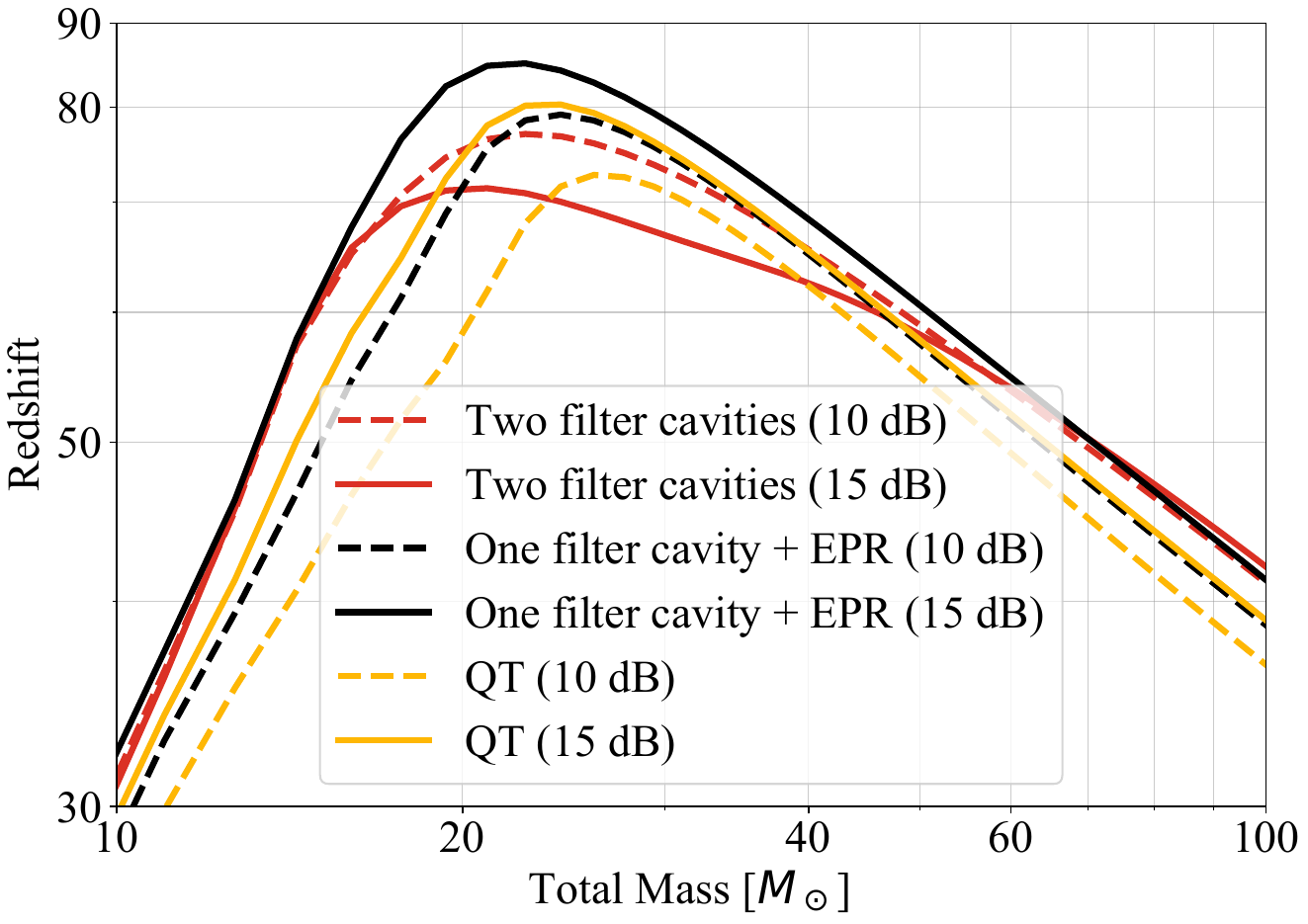}
    \caption{This figure displays the astrophysical reach of ET (three sets of interferometers with opening angles at 60 degrees) using different FDS schemes for equal-mass compact binaries, as a function of total source-frame mass} 
    \label{fig-horizon}
\end{figure}

Finally, in Fig.\,\ref{fig-horizon}, we show the horizon reach of different FDS schemes to the inspiral of equal-mass compact binaries as a function of total source-frame mass. Both the EPR and QT schemes give overall horizon improvement by increasing the squeezing level. The two filter cavity scheme results in less horizon reach mainly in the range of 20-40 solar masses after increasing the squeezing from 10 to 15\,dB. Compared with the two filter cavity scheme (10 dB squeezing), the EPR squeezing scheme (15 dB squeezing) expands the horizon reach of ET by a maximum of $\sim$ 10\% in the range of 15-60 solar masses.

\section{Conclusion}
The low-frequency detector of the Einstein telescope will operate at the detuned signal recycling regime, therefore it requires at least one more filter cavity for FDS compared with the the broadband detector, such as advanced LIGO/Virgo and Cosmic Explorer. As an underground detector, the space is precious. In this paper, we explored other approaches of FDS for the low-frequency detector of ET aiming for both saving the cost and potential enhancement of the sensitivity. Three alternative schemes are studied: (1) using one coupled filter cavity; (2) using EPR squeezing to replace two filter cavities by the interferometer itself; (3) using EPR squeezing to replace one filter cavity by the interferometer itself. We prove the first option is equivalent to the two filter cavities in theory, however, only the third option can be valid in practice. 
The second option is invalid because of the constraints from the predefined parameters of the interferometer.
In the third option, we opt to replace the filter cavity with a narrower bandwidth. 
It turns out, even though there exists $\sim$ 3dB inherent penalty and twice optical losses from the two beams, this design still shows sensitivity improvement benefiting from the lower effective loss from the longer effective filter cavity and higher input squeezing level, in particular below 10 Hz. In contrast, the higher input squeezing will degrade the sensitivity around 8\,Hz in the case of using two 1\,km filter cavities.
From an astrophysical perspective, this sensitivity enhancement enables the detection of gravitational waves emitted by more distant compact binaries. Notably, the most significant improvement is observed for binary stars with total source-frame mass ranging from 15 to 60 solar mass. 

The modern laser interferometric detector is a complex instrument, the optical losses could in principle arise from more sources than a single filter cavity, for example, the cavity mode mismatch and the optical path distortion from thermal effect. Here, indeed, we take into account 1000\,ppm SRC loss and 45\,ppm arm cavity loss in the EPR scheme. The noise contamination mechanism can be more, but modestly complex in reality, thanks to the low power operation in the ET low-frequency detector. In conclusion, the study in this work paves the way for a future implementation/upgrade of FDS in the low frequency detector of ET featuring EPR squeezing.

\begin{acknowledgments}
We would like to thank Yohei Nishino for providing the simulation code to produce the sensitivity in\, \cite{nishino2024frequency}, Haixing Miao and the members of the Institute for Gravitational Wave Astronomy at the University of Birmingham for fruitful discussions. T. Z., D. M., and X. P. acknowledge the support from the Institute for Gravitational Wave Astronomy. X. P. and Z. Z. have been supported by the School of Physical Science and Technology at Wuhan University. 
\end{acknowledgments}

\appendix
\section{Representation Transformation\label{app-loss-shotrdp}}
In the two-photon formalism, the field is defined with two orthogonal quadrature, \textit{i.e.} $\hat{\alpha}=[\hat{\alpha}_c,\hat{\alpha}_s]^{\rm T}$, where $\hat{\alpha}_c$, $\hat{\alpha}_s$ represent the amplitude and phase quadrature, respectively. The representation of a field in the sideband picture can be transformed into the quadrature field following the below relation:
\begin{equation}
    \begin{bmatrix}
        \hat{\alpha}_{\rm c}\\ \hat{\alpha}_{\rm s}
    \end{bmatrix}= \mathbb{M}\begin{bmatrix}
        \hat{\alpha}_{+}\\ \hat{\alpha}_-^{\dagger}
    \end{bmatrix},\quad \mathbb{M}= \frac{1}{\sqrt{2}}\begin{bmatrix}
        1&1\\-i&i
    \end{bmatrix} \, ,
\end{equation}
where $\hat{\alpha}_{+}$, $\hat{\alpha}^{\dagger}_-$ are the annihilation and creation operators, representing the upper sideband and lower sideband.

\section{Equivalence of Two filter cavities and coupled-cavity\label{app-2filter}}
The transfer function from the upper sideband of the input field  ($\hat{i}$) to the upper sideband of the output field ($\hat{o}$) in two filter cavities is:
    \begin{equation}
\hat{o}_{+}=f(\Omega)\hat{i}_{+}\, ,
\end{equation}
where
\begin{equation}\label{app-2filter}
\begin{aligned}
      f(\Omega)&=\frac{e^{i\phi_{1}(\Omega)}-\sqrt{R_{1}}}{1-\sqrt{R_{1}}e^{i\phi_{1}(\Omega)}}\frac{e^{i\phi_{2}(\Omega)}-\sqrt{R_{2}}}{1-\sqrt{R_{2}}e^{i\phi_{2}(\Omega)}}\, .\\
\end{aligned}
\end{equation}
Here $R_{1,2}$ is the amplitude reflectivity of the IM1 and IM2 of the two filter cavities, repectively.

\begin{figure}[!t]
    \centering
    \includegraphics[width=0.8\columnwidth]{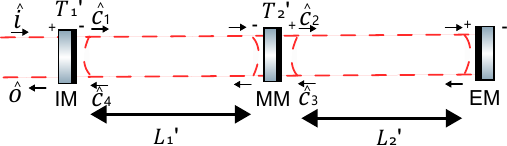}
    \caption{The figure shows the related fields in the coupled-cavity}
    \label{fig-coupled fields}
\end{figure}

As shown in Fig.\,\ref{fig-coupled fields}, the equations of the upper sideband of the related fields in the coupled-cavity are:
\begin{subequations}\label{eq-coupling1}
    \begin{equation}
\hat{o}_{+}=\sqrt{R'_1}\hat{i}_{+}+\sqrt{T'_1}\hat{c}_{4,+}  \quad ,
    \end{equation}
    
    \begin{equation}
\hat{c}_{1,+}=\sqrt{T'_1}\hat{i}_{+}-\sqrt{R'_1}\hat{c}_{4,+} \quad ,
    \end{equation}
    
\begin{equation}
    \hat{c}_{2,+} =\sqrt{T'_2}e^{i\phi'_1(\Omega)/2}\hat{c}_{1,+}+\sqrt{R'_2}\hat{c}_{3,+}\quad ,
\end{equation}

\begin{equation}
    \hat{c}_{3,+} =e^{i\phi'_2(\Omega)}\hat{c}_{2,+}  \quad ,
\end{equation}

\begin{equation}
    \hat{c}_{4,+} =\sqrt{T'_2}e^{i\phi'_1(\Omega)/2}\hat{c}_{3,+}\quad .
\end{equation}
\end{subequations}
We can derive the input-output relation of the upper sideband in the coupled-cavity from Eq.\,\ref{eq-coupling1}:
\begin{widetext}
\begin{equation}\label{couplinginout}\begin{aligned}
\hat{o}_{+}&=f'(\Omega)\hat{i}_{+}\\
    &=\frac{\sqrt{R_1'}-\sqrt{R_2'}e^{i\phi_{1}'(\Omega)}-\sqrt{R_1'}e^{i\phi_{2}'(\Omega)}+e^{i(\phi_{1}'(\Omega)+\phi_{2}'(\Omega))}}{1-\sqrt{R_1'}e^{i\phi_{2}'(\Omega)}-\sqrt{R_1'R_2'}e^{i\phi_{1}'(\Omega)}+\sqrt{R_1'}e^{i(\phi_{1}'(\Omega)+\phi_{2}'(\Omega))}}\hat{i}_{+}\, ,
\end{aligned}
\end{equation}
\end{widetext}
where $R_1'$ is the amplitude reflectivity of the IM of the coupled-cavity, and $R_2'$ is the amplitude reflectivity of the MM of the coupled-cavity.

Given the negligible transmissivity of both the IM and 
the MM of the coupled-cavity, as well as the slight round-trip phase within the coupled-cavity, we can expand the exponential function and square root in Eq.\,\ref{eq-coupling1} into second-order as:
\begin{subequations}\label{approx}
  \begin{equation}
        \sqrt{R_{1,2}}=\sqrt{1-T_{1,2}}=1-\frac{1}{2}T_{1,2}-\frac{1}{8}T_{1,2}^2\, ,
\end{equation}  
\begin{equation}
     e^{i\phi_{1,2}(\Omega)}=1+i\phi_{1,2}(\Omega)-\frac{1}{2}\phi_{1,2}^2(\Omega)\, ,
\end{equation}
  \begin{equation}
        \sqrt{R_{1,2}'}=\sqrt{1-T_{1,2}'}=1-\frac{1}{2}T_{1,2}'-\frac{1}{8}T_{1,2}'^2\, ,
\end{equation}  
\begin{equation}
     e^{i\phi_{1,2}'(\Omega)}=1+i\phi_{1,2}'(\Omega)-\frac{1}{2}\phi_{1,2}'^2(\Omega)\, .
\end{equation}
\end{subequations}
By substituting Eq.\,\ref{approx} with the same term in Eq.\,\ref{app-2filter} and Eq.\,\ref{couplinginout}, while keeping the second-order term, we can obtain:
\begin{subequations}
\begin{equation}\label{eq-2filterapp1}
    f(\Omega)=\frac{\frac{1}{4}T_1T_2-\phi_{1}(\Omega)\phi_{2}(\Omega)+\frac{i}{2}(T_2\phi_{1}(\Omega)+T_1\phi_{2}(\Omega))}{\frac{1}{4}T_1T_2-\phi_{1}(\Omega)\phi_{2}(\Omega)-\frac{i}{2}(T_2\phi_{1}(\Omega)+T_1\phi_{2}(\Omega))}\, ,
\end{equation}
    \begin{equation}\label{eq-coupledapp1}
f'(\Omega)=\frac{(T_2')^2+T_2'-\phi_1'(\Omega)\phi_2'(\Omega)+\frac{i}{2}(T_1'\phi_2'(\Omega)+T_2'\phi_1'(\Omega))}{(T_2')^2+T_2'-\phi_1'(\Omega)\phi_2'(\Omega)-\frac{i}{2}(T_1'\phi_2'(\Omega)+T_2'\phi_1'(\Omega))}\, .
    \end{equation}
\end{subequations}
Considering the following relations:
    \begin{equation}
    (T_2')^2\ll T_2'\ll T_1'\, ,
\end{equation}
we can eliminate the terms $(T_2')^2$ and $T_2'\phi_1'(\Omega)$ both in numerator and denominator of Eq.\,\ref{eq-coupledapp1}, resulting in a simplified expression as:
\begin{equation}\label{couplapp2}
f'(\Omega)=\frac{T_2'-\phi_{1}'(\Omega)\phi_{2}'(\Omega)+\frac{i}{2}T_1'\phi_{2}'(\Omega)}{T_2'-\phi_{1}'(\Omega)\phi_{2}'(\Omega)-\frac{i}{2}T_1'\phi_{2}'(\Omega)}\, .
\end{equation}

To replace the two filter cavities with the coupled-cavity, it necessitates:
\begin{equation}\label{eq-equvilence}
f(\Omega)=f'(\Omega)\, .
\end{equation}
By comparing the real part and imaginary part of either the denominator or numerator of Eq.\,\ref{eq-2filterapp1} and Eq.\,\ref{couplapp2}, and setting them identical, we can derive the following equations from Eq.\,\ref{eq-equvilence}:
\begin{subequations}\label{2=c}
    \begin{equation}
       T_2'-\phi_{1}'(\Omega)\phi_{2}'(\Omega)= \frac{1}{4}T_1T_2-\phi_{1}(\Omega)\phi_{2}(\Omega)\, ,
    \end{equation} \begin{equation}T_1'\phi_{2}'(\Omega)= T_2\phi_{1}(\Omega)+T_1\phi_{2}(\Omega)\, .
    \end{equation}
\end{subequations}
By bringing 
\begin{subequations}
    \begin{equation}
    \phi_{1,2}(\Omega)=2(\delta\omega_{1,2}+\Omega)\tau_{1,2},\,T_{1,2}=4\gamma_{1,2}\tau_{1,2}\, ,
\end{equation}
\begin{equation}
    \phi_{1,2}'(\Omega)=2(\delta\omega_{1,2}'+\Omega)\tau_{1,2}'\, ,
\end{equation}
\begin{equation}
    T_1'=4\gamma_1'\tau_1',\,T_2'=4\omega_s^2\tau_1'\tau_2'\, ,
\end{equation}
\end{subequations}
into Eq.\ref{2=c}, where $\tau_{1,2}=\frac{L_{1,2}}{c}$ is the single trip time the light acquires in the corresponding cavity of two filter cavities, and $\tau_{1,2}'=\frac{L_{1,2}'}{c}$ is the single trip time the light acquires in the corresponding cavity of the coupled-cavity, we can rewrite Eq.\,\ref{2=c} into the polynomial of sideband frequency $\Omega$ as:
\begin{subequations}\label{eq-Omega}
    \begin{equation}
    \begin{aligned}
&\tau_1'\tau_2'[\Omega^2+(\delta\omega_1'+\delta\omega_2')\Omega+(\delta\omega_1'\delta\omega_2'-\omega_{s}^2)]=\\
&\tau_1\tau_2[\Omega^2+(\delta\omega_1+\delta\omega_2)\Omega+(\delta\omega_1\delta\omega_2-\gamma_1\gamma_2)]\, ,
\end{aligned}
\end{equation}
\begin{equation}
\begin{aligned}
\tau_1'\tau_2'[\gamma_1'\Omega+\gamma_1'\delta\omega_2']=\tau_1\tau_2[(\gamma_1+\gamma_2)\Omega+(\gamma_1\delta\omega_2+\gamma_2\delta\omega_1)]\, .
\end{aligned}
\end{equation}
\end{subequations}

Hence, to fulfill Eq.\,\ref{eq-Omega} for any sideband frequency $\Omega$, it evidently necessitates the coefficients of each order of $\Omega$ to be the same:
\begin{subequations}\label{app-2}
\begin{equation}
\tau_1'\tau_2'=\tau_1\tau_2\, ,
\end{equation}
\begin{equation}\gamma_1'=\gamma_1+\gamma_2\, ,
\end{equation}
\begin{equation}
\gamma_1'\delta\omega_2'=\gamma_1\delta\omega_2+\gamma_2\delta\omega_1\, ,
\end{equation}
\begin{equation}
\delta\omega_1'+\delta\omega_2'=\delta\omega_1+\delta\omega_2\, ,
\end{equation}
\begin{equation}
    \omega_{s}^2-\delta\omega_1'\delta\omega_2'=\gamma_1\gamma_2-\delta\omega_1\delta\omega_2\, .
\end{equation}
\end{subequations}
We can derive Eq.\,\ref{2fltertocoupling} from Eq.\,\ref{app-2}.

In the following discussion, we'll prove the equivalence between the SRC-arm coupled-cavity and a coupled-cavity shown in (b) of Fig.\ref{fig-2filter} to achieve the proper rotation for the idler beam. The frequency of the idler field in the SRC-arm coupled-cavity is given by:
\begin{subequations}\label{app-3}
    \begin{equation}\label{app-SRCphase}
\phi_{1}'(\Omega)\to\phi_{\text{SRC}}(\Omega)=2(\delta\omega_{\text{SRC}}+\Delta+\Omega)\tau_{\text{SRC}}\, ,
    \end{equation}
    \begin{equation}\phi_{2}'(\Omega)\to\phi_{\text{arm}}(\Omega)=2(\Delta+\Omega)\tau_{\text{arm}}\, ,
    \end{equation}
\end{subequations}
where $\tau_{\text{SRC}}=\frac{L_{\rm SRC}}{c}$ is the single trip time of the SRC, $\phi_{\rm arm}(\Omega)$ is the round trip phase the idler beam acquires in the arm cavity, $\tau_{\text{arm}}=\frac{L_{\rm arm}}{c}$ is the single trip time of the arm cavity. 


If the SRC-arm coupled-cavity is equivalent to the coupled-cavity for the idler beam, it requires:
\begin{equation}\label{app-13}
f_{\rm SRC-arm}(\Omega)=f'(\Omega)\, ,
\end{equation}
where
\begin{equation}\label{app-eq14}
    f_{\rm SRC-arm}(\Omega)=\frac{T_{\text{ITM}}-\phi_{\text{SRC}}(\Omega)\phi_{\text{arm}}(\Omega)+\frac{i}{2}T_{\text{SRM}}\phi_{\text{arm}}(\Omega)}{T_{\text{ITM}}-\phi_{\text{SRC}}(\Omega)\phi_{\text{arm}}(\Omega)-\frac{i}{2}T_{\text{SRM}}\phi_{\text{arm}}(\Omega)}\, .
\end{equation}

Applying a similar methodology as used in deriving Eq.\,\ref{app-2}, we could obtain the following parameter equations by bringing Eq.\,\ref{app-eq14} and Eq.\ref{eq-coupledapp1} into the Eq.\,\ref{app-13}:
\begin{subequations}\label{app-5}
    \begin{equation}
\tau_{\text{SRC}}\tau_{\text{arm}}=\tau_1'\tau_2' \, ,
    \end{equation}
\begin{equation}\gamma_{\text{SRC}}=\gamma_1'\, ,
    \end{equation}\begin{equation}\Delta=\delta\omega_2'\, ,
    \end{equation}
    \begin{equation}
        (\omega_s^{\text{SRC-arm}})^2-(\delta\omega_{\text{SRC}}+\Delta)\Delta=(\omega_s)^2-\delta\omega_1'\delta\omega_2'\, ,
    \end{equation}
    \begin{equation}
\delta\omega_{\text{SRC}}+2\Delta=\delta\omega_1'+\delta\omega_2'\, .
    \end{equation}
\end{subequations}
We can derive Eq.\,\ref{eq-constraints} from Eq.\,\ref{app-5}.

\begin{figure}[!t]
\includegraphics[width=\columnwidth]{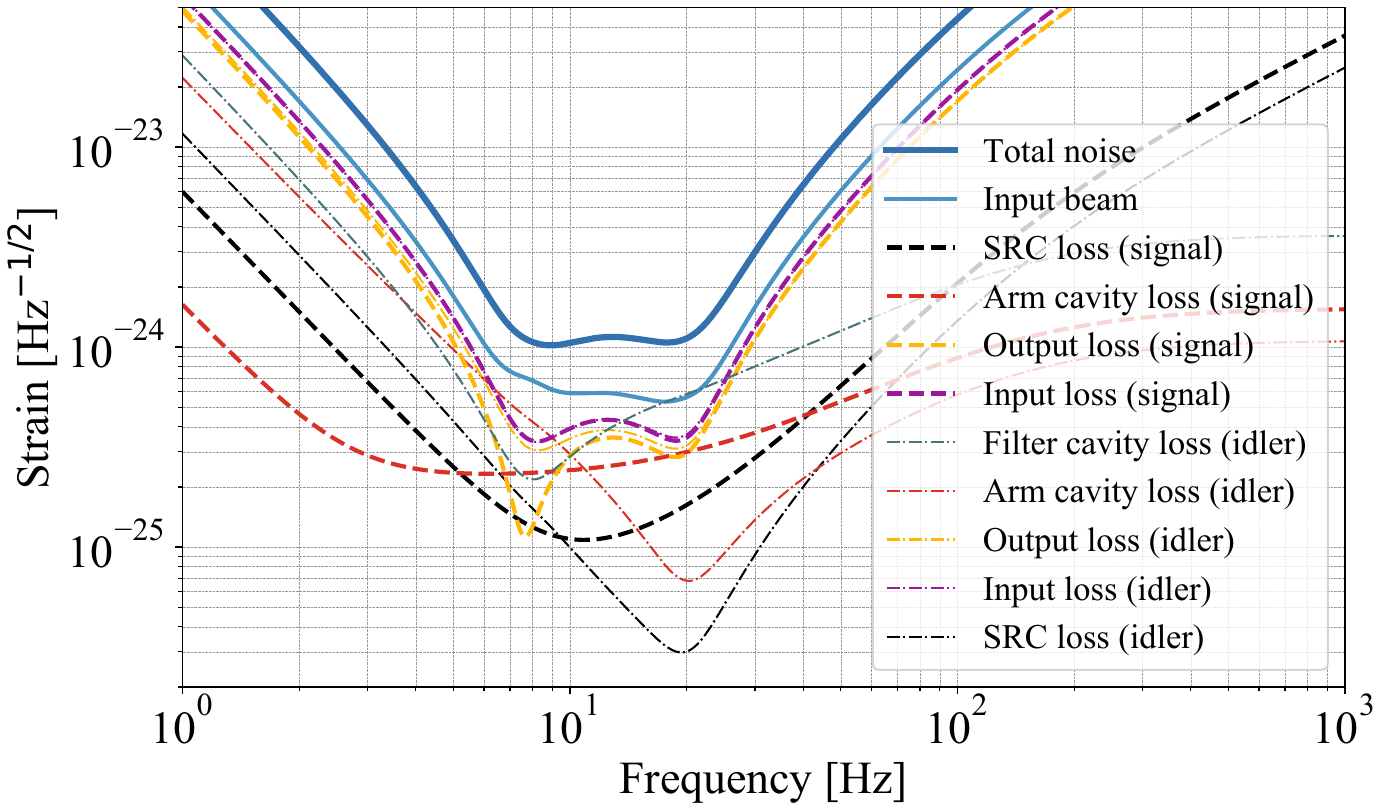}
    \caption{The figure shows the noise budget of the quantum noise using the EPR squeezing scheme (15 dB squeezing). The input and output losses of both beams are significant below 6 Hz and above 15 Hz, the arm cavity loss of the idler beam is dominant between 6 Hz and 9 Hz, and the filter cavity loss of the idler beam is significant from 15 Hz to 25 Hz}
    \label{fig-EPR-2filter}
\end{figure}

\section{Input output relations}\label{app-loss}
\begin{figure}[!t]
    \centering
\includegraphics[width=\columnwidth]{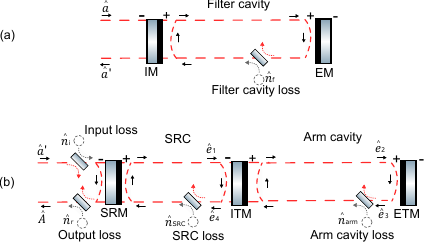}
    \caption{(a) The upper figure shows how the filter cavity loss couples with the beams when beams go through the filter cavity.
(b) The lower figure shows how the input loss, SRC loss, arm cavity loss and output loss couple with the beams when beams go through the interferometer}
    \label{fig-SRCloss}
\end{figure}

As shown in Fig.\,\ref{fig-SRCloss}, the output fields not only contain the contribution from the input fields but also the contribution from the optical losses and the GW signal, as:
\begin{subequations}\label{app-sensitivity}
    \begin{equation}
    \begin{aligned}
     \hat{A}=&\sqrt{1-\epsilon_{\rm r}^2} [\mathbb{A}\hat{a}+\mathbb{A}\sqrt{\epsilon_{\rm i}}\hat{n}_{\rm i}+\mathbb{A}_{\rm f}\hat{n}_{\rm f}+\mathbb{A}_{\rm SRC}\hat{n}_{\rm SRC}\\
&+\mathbb{A}_{\rm arm}\hat{n}_{\rm arm}]+\epsilon_{\rm r}\hat{n}_{\rm r}+\mathbf{R}\hat{x}  \, ,
    \end{aligned}
\end{equation}
        \begin{equation}
    \begin{aligned}
     \hat{B}=&\sqrt{1-\epsilon_{\rm r}^2} [\mathbb{B}\hat{b}+\mathbb{B}\sqrt{\epsilon_{\rm i}}\hat{l}_{\rm i}+\mathbb{B}_{\rm f}\hat{l}_{\rm f}+\mathbb{B}_{\rm SRC}\hat{l}_{\rm SRC}\\&+\mathbb{B}_{\rm arm}\hat{l}_{\rm arm}]+\epsilon_{\rm r}\hat{l}_{\rm r} \, ,
    \end{aligned}
\end{equation}
\end{subequations}
where $\hat{n}$ and $\hat{l}$ are from the optical losses of the signal beam and idler beam, $\mathbb{A}$ and $\mathbb{B}$ are the transfer matrices from the corresponding fields to the output fields, $\epsilon$ are the loss coefficients, and the subscript "f" stands for filter cavity loss, 
the subscript "i" stands for the input loss, the subscript "SRC" stands for the SRC loss, the subscript "arm" stands for the arm cavity loss, and the subscript "r" stands for the readout loss.

\subsection{Derivitation of $\mathbb{A}_{\rm arm}$\label{app-Aarm}}
The matrix $\mathbb{A}_{\rm arm}$ contains the contribution from the shot noise and radiation pressure noise of the arm cavity loss as:
\begin{equation}
    \mathbb{A}_{\rm arm}=\sqrt{\epsilon_{\rm arm}}(\mathbb{S}_{\rm arm}+\mathbb{F}_{\rm arm})\, ,
\end{equation}
where the matrix $\mathbb{S}_{\rm arm}$ represents the contribution from the shot noise of the arm cavity loss, $\mathbb{F}_{\rm arm}$ represents the contribution from the radiation pressure noise of the arm cavity loss. The formula of $\mathbb{S}_{\rm arm}$ is:
    \begin{equation}
    \mathbb{S}_{\rm arm}=\mathbb{M}\begin{bmatrix}
        f_{\hat{e}_{3} \to \hat{A}}(\Omega) &0\\0&f_{\hat{e}_{3} \to \hat{A}}^{\dag}(-\Omega)
    \end{bmatrix}\mathbb{M}^{-1}\, ,
\end{equation}
where
    \begin{equation}
    f_{\hat{e}_{3} \to \hat{A}}(\Omega)=\frac{\sqrt{T_{\text{SRM}}}\mathcal{T}'_{\text{arm}}(\Omega)e^{i\phi'_{\text{SRC}}(\Omega)/2}}{1-\sqrt{R_{\text{SRM}}}\mathcal{R}'_{\text{arm}}(\Omega)e^{i\phi'_{\text{SRC}}(\Omega)}}\, .
\end{equation}
Here
\begin{subequations}\label{app-armreflecttrans}
    \begin{equation}\label{app-armreflect}
    \mathcal{R}'_{\text{arm}}(\Omega)=-\sqrt{R_{\text{ITM}}}+\frac{T_{\text{ITM}}e^{i\phi'_{\text{arm}}(\Omega)}}{1-\sqrt{R_{\text{ITM}}}e^{i\phi'_{\text{arm}}(\Omega)}}\, ,
\end{equation}
\begin{equation}\label{app-armtrans}
\mathcal{T}'_{\text{arm}}(\Omega)=\frac{\sqrt{T_{\text{ITM}}}e^{i\phi'_{\text{arm}}(\Omega)/2}}{1-\sqrt{R_{\text{ITM}}}e^{i\phi'_{\text{arm}}(\Omega)}}\, ,
\end{equation}
\end{subequations}
are the amplitude reflectivity and transmissivity of the arm cavity when the signal beam is injected from the ITM, and 
\begin{equation}
    \phi'_{\rm SRC}(\Omega)=\varphi_{\rm SRC}+2\Omega\frac{L_{\rm SRC}}{c}\, , \phi'_{\rm arm}(\Omega)=2\Omega\frac{L_{\rm arm}}{c} \, ,
\end{equation}
are the round trip phase of the upper sideband of the signal beam acquired in the corresponding cavity.

The formula of $\mathbb{F}_{\rm arm}$ is:
    \begin{equation}
\mathbb{F}_{\rm arm}=\mathbf{R}\chi_{xx}\textbf{F}_{\rm arm}\, ,
\end{equation}
where
$\chi_{xx}$ is the transfer function from the force on the test mirror to the differential displacement of the test mass:
\begin{equation}
    \chi_{xx}=M\Omega^2\, .
\end{equation}

The formula of $\mathbf{F}_{\rm arm}$ is:
\begin{equation}
\mathbf{F}_{\rm arm}=2\hbar k\mathbf{E}\mathbb{M}\begin{bmatrix}
        f_{\hat{e}_{3} \to \hat{e}_{2}}(\Omega)&0\\0&f_{\hat{e}_{3} \to \hat{e}_{2}}^{\dag}(-\Omega)
    \end{bmatrix}\mathbb{M}^{-1}\, ,
\end{equation}
where $\hbar$ is the reduced Plank constant, $k=\frac{\omega_0}{c}$ is the wave number of the signal beam, $\omega_0$ is the frequency of the carrier beam, and $f_{\hat{e}_{3} \to \hat{e}_{2}}(\Omega)$ is:
\begin{equation}
f_{\hat{e}_{3} \to \hat{e}_{2}}(\Omega)=
\frac{\mathcal{R}_{\text{SRC}}'(\Omega)e^{i\phi_{\text{arm}}'(\Omega)}}{1-\mathcal{R}_{\text{SRC}}'(\Omega)e^{i\phi_{\text{arm}}'(\Omega)}}\, .
\end{equation}
Here $\mathbf{E}$ is the classical electric field in the arm cavity:
\begin{equation}
    \mathbf{E}=\begin{bmatrix}
        2E&0
    \end{bmatrix}\, ,
\end{equation}
and $E$ is the amplitude of the electric field:
\begin{equation}
    E=\sqrt{\frac{cI_0}{2\hbar\omega_0 L_{\rm arm}}}\, ,
\end{equation}
where $I_0$ is the power of the laser at the beam splitter.
The radiation pressure force from the arm cavity loss is:
 \begin{equation}
     \hat{F}_{\rm arm}=\mathbf{F}_{\rm arm}\sqrt{\epsilon_{\rm arm}}\hat{n}_{\rm arm}\, .
 \end{equation}

The formula of $\mathbf{R}$ is:
\begin{equation}
    \mathbf{R}=2ik E\mathbb{S}_{\rm arm}\begin{bmatrix}
         0\\1
    \end{bmatrix}\, .
\end{equation}

\subsection{Derivitation of $\mathbb{A}_{\rm SRC}$}
Similarly, the matrix $\mathbb{A}_{\rm SRC}$ contains the contribution from the
shot noise and radiation pressure noise of the SRC loss as:
\begin{equation}
    \mathbb{A}_{\rm SRC}=\sqrt{\epsilon_{\rm SRC}}(\mathbb{S}_{\rm SRC}+\mathbb{F}_{\rm SRC})\, .
\end{equation}
To derive the $\mathbb{S}_{\rm SRC}$ and $\mathbb{F}_{\rm SRC}$, it requires the transfer function from the field $\hat{e}_4$ to the output $\hat{A}$ and to the field $\hat{e}_2$. In the sideband picture, they are:
\begin{subequations}
\begin{equation}
   f_{\hat{e}_4 \to \hat{A}}(\Omega)=\frac{\sqrt{T_{\text{SRM}}}e^{i\phi'_{\text{SRC}}(\Omega)/2}}{1-\sqrt{R_{\text{SRM}}}\mathcal{R}'_{\text{arm}}(\Omega)e^{i\phi'_{\text{SRC}}(\Omega)}}\, ,
\end{equation} 
\begin{equation}
\begin{aligned}
    f_{\hat{e}_4 \to \hat{e}_2}(\Omega)=&
\frac{\sqrt{R_{\text{SRM}}}e^{i\phi_{\text{SRC}}'(\Omega)}}{1-\sqrt{R_{\text{SRM}}}\mathcal{R}_{\text{arm}}'(\Omega)e^{i\phi_{\text{SRC}}'(\Omega)}}\\&\times\frac{\sqrt{T_{\text{ITM}}}e^{i\phi_{\text{arm}}'(\Omega)/2}}{1-\sqrt{R_{\text{ITM}}}e^{i\phi_{\text{arm}}'(\Omega)}}\, .
\end{aligned}
\end{equation}
\end{subequations}
The following calculations are similar to those in Appendix.\,\ref{app-Aarm}.

\subsection{Derivitation of $\mathbb{A}$ and $\mathbb{A}_{\rm f}$}
Similarly, the matrix $\mathbb{A}$ and $\mathbb{A}_{\rm f}$ contain the contribution
from the shot noise and radiation pressure noise of the input loss and filter cavity loss as:
\begin{subequations}
    \begin{equation}
    \mathbb{A}=(\mathbb{S}_{\rm i}+\mathbb{F}_{\rm i})\mathbb{M}\begin{bmatrix}
          \mathcal{R}'_{\rm f}(\Omega) &0\\0&\mathcal{R}_{\rm f}^{'\dag}(-\Omega)
     \end{bmatrix}\mathbb{M}^{-1}\, ,
\end{equation}
\begin{equation}
    \mathbb{A}_{\rm f}=\sqrt{\epsilon_{\rm f}}(\mathbb{S}_{\rm i}+\mathbb{F}_{\rm i})\mathbb{M}\begin{bmatrix}
         \mathcal{T}'_{\rm f}(\Omega) &0\\0&  \mathcal{T}_{\rm f}^{'\dag}(-\Omega)
     \end{bmatrix}\mathbb{M}^{-1}\, ,
\end{equation}
\end{subequations}
where 
\begin{subequations}
\begin{equation}
        \mathcal{R}'_{\rm f}(\Omega)=-\sqrt{R_{1}}+\frac{T_{1}e^{i\phi'_{\rm f}(\Omega)}}{1-\sqrt{R_{1}}e^{i\phi'_{\rm f}(\Omega)}}\, ,
\end{equation}

\begin{equation}
        \mathcal{T}'_{\rm f}(\Omega)=\frac{\sqrt{T_{1}}e^{i\phi'_{\rm f}(\Omega)/2}}{1-\sqrt{R_{1}}e^{i\phi'_{\rm f}(\Omega)}}\, ,
\end{equation}
\end{subequations}
is the effective reflectivity and transmissivity of the filter cavity when the signal beam is injected from the IM of the filter cavity. Here
\begin{equation}
    \phi'_{\rm f}(\Omega)=2\Omega\frac{L_1}{c}\, ,
\end{equation}
is the round-trip phase of the signal beam acquired in the filter cavity.

To derive the $\mathbb{S}_{\rm i}$ and $\mathbb{F}_{\rm i}$, it requires the transfer function from the field $\hat{a}'$ to the output $\hat{A}$ and to the field $\hat{e}_2$. In the sideband picture, they are:
\begin{subequations}
\begin{equation}
   f_{\hat{a}' \to \hat{A}}(\Omega)=-\sqrt{R_{\text{SRM}}}+\frac{\mathcal{R}'_{\text{arm}}(\Omega)T_{\text{SRM}}e^{i\phi'_{\text{SRC}}(\Omega)}}{1-\sqrt{R_{\text{SRM}}}\mathcal{R}'_{\text{arm}}(\Omega)e^{i\phi'_{\text{SRC}}(\Omega)}}\, ,
\end{equation} 
\begin{equation}
\begin{aligned}
    f_{\hat{a}' \to \rm \hat{e}_2}(\Omega)=&
\frac{\sqrt{T_{\text{SRM}}}e^{i\phi_{\text{SRC}}'(\Omega)/2}}{1-\sqrt{R_{\text{SRM}}}\mathcal{R}_{\text{arm}}'(\Omega)e^{i\phi_{\text{SRC}}'(\Omega)}}\\&\times\frac{\sqrt{T_{\text{ITM}}}e^{i\phi_{\text{arm}}'(\Omega)/2}}{1-\sqrt{R_{\text{ITM}}}e^{i\phi_{\text{arm}}'(\Omega)}}\, .
\end{aligned}
\end{equation}
\end{subequations}
The following calculations are similar to those in Appendix.\,\ref{app-Aarm}.

\subsection{Derivitation of $\mathbb{B}$, $\mathbb{B}_{\rm f}$, $\mathbb{B}_{\rm SRC}$, and $\mathbb{B}_{\rm arm}$}
For the idler beam, there is no contribution from the radiation pressure force to the matrix $\mathbb{B}$, $\mathbb{B}_{\rm f}$, $\mathbb{B}_{\rm SRC}$, and $\mathbb{B}_{\rm arm}$, \textit{i.e.} $\mathbb{F}_{\rm i, \,f, \,SRC, \,arm}=0$. To derive these matrices, we only need to substitute the round-trip phase of the signal beam with the round-trip phase of the idler beam, as:
\begin{subequations}
    \begin{equation}
    \phi'_{\rm f}(\Omega)\longrightarrow\phi_{\rm f}(\Omega)=2(\Delta+\Omega)\frac{L_1}{c}\, ,
\end{equation}
    \begin{equation}
    \phi'_{\rm SRC}(\Omega)\longrightarrow\phi_{\rm SRC}(\Omega)\, ,
\end{equation}
    \begin{equation}
    \phi'_{\rm arm}(\Omega)\longrightarrow\phi_{\rm arm}(\Omega)\, .
\end{equation}
\end{subequations}


\bibliography{references}

\end{document}